# Individual participant data from digital sources informed and improved precision in the evaluation of predictive biomarkers in Bayesian network meta-analysis.

Chinyereugo M Umemneku-Chikere[1]*, Lorna Wheaton[1],
Heather Poad[1], Devleena Ray[1], Ilse Cuevas Andrade[1], Sam Khan[2],
Paul Tappenden[3], Keith R Abrams[4,5], Rhiannon K Owen[6], Sylwia Bujkiewicz[1]

1. Biostatistics Research Group, Department of Population Health Sciences, University of Leicester, Leicester, UK.
2. Leicester Cancer Research Centre, Robert Kilpatrick Clinical Sciences Building, University of Leicester, Leicester, UK.
3. School of Health and Related Research, University of Sheffield, Sheffield, UK.
4. Department of Statistics, University of Warwick, Coventry, UK.
5. Centre for Health Economics, University of York, York, UK.
6. Swansea University Medical School, Swansea University, Swansea, UK.

*Corresponding Author,* Email: chinyereugo.umemneku-chikere@york.ac.uk

**Abstract**:

**Objective:** We aimed to develop a meta-analytic model for evaluation of predictive biomarkers and targeted therapies, utilising data from digital sources when individual participant data (IPD) from randomised controlled trials (RCTs) are unavailable.

**Design and Setting**: A Bayesian network meta-regression model, combining aggregate data (AD) from RCTs and IPD, was developed for modelling time-to-event data to evaluate predictive biomarkers. IPD were sourced from electronic health records, using target trial emulation approach, or digitised Kaplan-Meier curves. The model is illustrated using two examples; breast cancer with a hormone receptor biomarker, and metastatic colorectal cancer with the Kirsten Rat Sarcoma (KRAS) biomarker.

**Results**: The model developed allowed for estimation of treatment effects in two subgroups of patients defined by their biomarker status. Effectiveness of taxane did not differ in hormone receptor positive and negative breast cancer patients. Epidermal growth factor receptor (EGFR) inhibitors were more effective than chemotherapy in KRAS wild type colorectal cancer patients but not in patients with KRAS mutant status. Use of IPD reduced uncertainty of the sub-group specific treatment effect estimates by up to 49%.

**Conclusion**: Utilisation of IPD allowed for more detailed evaluation of predictive biomarkers and cancer therapies and improved precision of the estimates compared to use of AD alone.

**Keywords (6)**: IPD network meta-analysis, network meta-regression, predictive biomarker

## 1. Introduction

Predictive biomarkers and associated targeted therapies are at the centre of precision medicine and, therefore, are of great interest to patients, researchers, pharmaceutical companies, regulators, such as the European Medicines Agency (EMA) or Food and Drug Administration (FDA), and health technology assessment agencies such as the National Institute for Health and Care Excellence (NICE). When new cancer therapies are developed, they may be more effective in patients who harbour a specific biomarker. This may in turn lead to greater gains for patients in terms of overall survival and/or health-related quality of life. For example, the aromatase inhibitors (AI) like letrozole and anastrozole have shown to be effective treatments in breast cancer patients who are hormone receptor positive [1, 2]. Anaplastic lymphoma kinase (ALK) inhibitors such as Crizotinib or Alectanib have also shown to be effective in Non-Small Cell Lung Carcinomas (NSCLC) patients who harbour echinoderm microtubule-associated protein-like 4 (EML4) and ALK rearrangements. Crizotinib is also an effective therapy in stage IV lung cancer patients with c-ros oncogene 1 (ROS1) rearrangement [3].

Randomised controlled trials (RCTs) are considered the 'gold standard' approach when evaluating treatment efficacy, which may include evaluation of effectiveness within subgroups of patients [4]. However, there are limitations; for example, due to cost, safety, or ethical reasons, in undertaking RCTs in order to provide head-to-head comparisons of all the therapies administered for a target disease such as the human epidermal receptor 2 (HER2) positive advanced breast cancer (ABC). Information from multiple RCTs can be synthesised using network meta-analysis (NMA) [5-7] to obtain estimates of relative effectiveness of all competing therapies based on direct and indirect comparisons.

In situations where RCTs reported the treatment effects within subgroups according to the biomarker status, such as hormone receptor positive or negative subgroups, a NMA can be undertaken within each subgroup. However, if the subgroup analysis is not reported in the RCTs, but the proportion of patients recruited within each subgroup is known, a network meta-regression (NMR) could be considered [8, 9]. In NMR, the proportion of biomarker positive or negative patients are fitted as a covariate in the model to estimate the relative treatment effects within these subgroups. However, such aggregate level information is limited and individual participant data (IPD), when available, may provide significantly better quality of information used in an IPD NMA, using either a one-stage or two-stage approach [10] When IPD are available from a subset of studies, they can be combined with aggregate data (AD) to estimate the treatment effects [11, 12].

Proctor et al[13] developed a NMR method for combining evidence from IPD and AD to estimate the indirect treatment effect in binary biomarker subgroups using a binary response from published RCTs. Using a binary outcome, it was straightforward to convert AD to IPD for studies where all patients were biomarker positive and the proportion of responders was given. This is more complex for other outcomes and in situations when ADs are available from trials in patients of mixed biomarker status when the use of aggregate data alone will likely introduce bias. In such circumstances, IPD may provide great benefit.

In this paper, we build on the method by Proctor et al [13]. to develop a one-stage Bayesian NMA combining IPD and AD for estimation of treatment effects on time-to-event outcomes within subgroups of patients harbouring a specific biomarker type. In the absence of direct access to IPD from RCTs, we utilise other sources of IPD by either emulating trials from Electronic Health Records (EHR) or by digitizing Kaplan-Meier curves from RCTs. We illustrate the method using two clinical case studies; in breast cancer and metastatic colorectal cancer.

Section two of this paper discusses the illustrative examples (or case study) explored in this paper. Section three describes our model and how the IPD were obtained. Section four discusses the results from the applied case studies. Finally, the paper concludes in section five with a discussion highlighting the strengths, limitations, and recommendations for further research.

**2. Illustrative examples**

Two case studies were explored in this paper. The first example aims to estimate the relative effectiveness of therapies for advanced breast cancer (ABC) patients within hormone receptor positive and negative subgroups. We consider the effectiveness of taxanes used as monotherapy or in combination with chemotherapy. The hormone receptor positive (HR+ve) subgroup includes patients who are progesterone and/or oestrogen positive; while the hormone receptor negative (HR-ve) subgroup includes patients who are progesterone and oestrogen negative. In the second example, we aim to estimate the treatment effects of therapies targeted on Vascular Endothelial Growth Factor (VEGF) and epidermal growth factor receptor (EGFR) administered to patients with colorectal cancer within two patients' populations defined by the Kirsten rat sarcoma (KRAS) biomarker; those harbouring either wild type or mutant KRAS status.

## 3. Methods

In this section, we first discuss sources of data including methods for obtaining IPD; either from electronic health records (EHRs) or by digitizing Kaplan Meier curves. We then describe our one-stage NMA method for estimation of treatment effects within two biomarker subgroups simultaneously by synthesising a mixture of IPD and AD. At the end of this section, we briefly discuss how we apply the meta-analytic method to analyse different subsets of data to demonstrate the added value of IPD.

### 3.1. Sources of data

#### 3.1.1. Summary data from randomised controlled trials

In our example in advanced breast cancer, ADs were extracted from a systematic review carried out by Ghersi et al [14], who investigated the effectiveness of taxanes (X) used as monotherapy or in combination with chemotherapy (CX). Data were obtained from five trials comparing X with chemotherapy (C) and eight trials comparing CX with C alone.

Summary data for the example in metastatic colorectal cancer were obtained from the systematic review by Poad *et al* [15]. There were five RCTs that reported the efficacy of anti-vascular endothelial growth factor (anti-VEGF) therapies plus C (VEGF+C) vs C and ten RCTs evaluating anti-epidermal growth factor receptor (anti-EGFR) therapies plus C (EGFR+C) vs C. Chemotherapy treatments included capecitabine, FOLFOX (folinic acid, fluorouracil and oxaliplatin), CAPOX (capecitabine and oxaliplatin) and FOLFIRI (folinic acid, fluorouracil, and irinotecan hydrochloride). Therapies in the EGFR treatment class included cetuximab and panitumumab. The VEGF treatment class include bevacizumab, cediranib and sorafenib.

In both case studies data were obtained on the treatment effects on overall survival (OS) measured by hazard ratios (HRs) and the proportion of patients within each biomarker subgroup. Data for both case studies and relevant references are listed in the Appendix.

#### 3.1.2. Individual participant data

For the case study in advanced breast cancer, IPD were obtained from Simulacrum, which is a synthetic dataset imitating the EHRs held in the Systematic Anti-Cancer Treatment (SACT) data resource by the National Cancer Registration and Analysis Service (NCRAS) within National Health Service (NHS) Digital. Data set consisted of the total of 1370ndividuals who received one of the therapies under consideration (X, CX or C). To select patients and estimate treatment effect from the synthetic EHR the data were selected and analysed using the target trial emulation approach described below.

For the colorectal case study, IPD were obtained by digitizing Kaplan-Meier curves using the method by Guyot *et al* [16]. Kaplan-Meier curves were available for ten RCTs investigating the effectiveness of EGFR+C compared to C alone and for subgroups of patients defined by the KRAS status. The list of RCTs is provided in the Appendix.

### 3.2. Emulation of Target Trials

We utilised the Simulacrum data set to emulate a series of trials of CX vs C and X vs C in advanced breast cancer patients using a target trial approach [17]. In the first instance, we specified the key components of the target trial protocol, which (following recommendation by Hernan and Robins [17]) included eligibility criteria, treatment strategies, assignment procedures, the follow-up period, outcome, causal contrast, and statistical analysis. We aimed to emulate target trials that would complement the AD from the literature review and include the hormone receptor status at the individual level.

#### 3.2.1. Eligibility criteria

Patients aged 18 years and above with ABC, who received the treatment as their first lien therapy, were selected into the trials.

#### 3.2.2. Treatment strategies

Three treatment classes were considered; monotherapy X, combination therapy CX and the control treatment C. Taxanes included paclitaxel and docetaxel of any dose. Chemotherapy included fluorouracil, epirubicin, cyclophosphamide, vinorelbine, capecitabine, and carboplatin at any dose.

#### 3.2.3. Assignment procedures

Patients were matched to ensure that each trial had a similar number of patients in each arm and minimal discrepancies in the baseline characteristics (hormone receptor status and age) between arms. The matching was done using the propensity scoring approach via the MatchIt package in the RStudio [18] software. Given that the patients were not randomly allocated, it was assumed that no unmeasured confounding at baseline conditional on measured prognostic factors can influence the treatment effect. The prognostic factors used to match the patients in each study were age and hormone receptor biomarker status.

#### 3.2.4. The follow up period

The patients were followed from the initiation of first-line therapy until the start of second line therapy (where patients were censored), censoring or death.

### 3.2.5. Outcome and causal contrast

The outcome of interest was overall survival defined as time from treatment initiation to censoring or death. Baseline characteristics for each group were summarised to ensure that the covariates were similarly distributed across the treatment arms. We estimated the per-protocol effect in all emulated trials.

### 3.2.6. Statistical analysis

To estimate the treatment effects, a parametric Weibull proportional hazards regression model was employed in a Bayesian framework as described in Section 3.3.1. The model included a treatment-biomarker interaction term to allow for estimation of treatment effects in two biomarker subgroups.

## 3.3. Meta-analytic model

Building on the model by Proctor et al [13], we develop a NMA model that allows for simultaneous synthesis of data at either aggregate level or individual participant level. Information on the biomarker status is included directly from studies with IPD available and as the proportion of biomarker positive patients where only ADs were available. The model assumes that no subgroup analyses were reported for treatment effects within the biomarker groups, but such information can be easily incorporated in the model by treating the subgroups as individual studies with proportions of biomarker positive equal to one or zero. The first part of the model describes the contribution of the individual level data to the model, the second part described the AD are modelled and this is followed by discussion of how the two parts of the model can be combined.

### 3.3.1. Part I: NMA model for IPD studies

To model the treatment effect at the individual participant level, we model time-to-event data assuming a Weibull distribution. This is a flexible distribution which reduces to an exponential distribution in the presence of constant hazard, but the model could be adapted easily assuming alternative distributions. For patient $i$ in study $j$, time to event follows a Weibull distribution with a shape parameter $\gamma_j$ and a scale parameter $\lambda_{ij}$

$$t_{ij} \sim Weibull\left(\gamma_j, \lambda_{ij}\right). \tag{1}$$

The log hazard ($\lambda_{ij}$), depends on the biomarker status $X_{ij}$ of the patient (with $X_{ij} = 0$ for biomarker negative patients and $X_{ij} = 1$ for biomarker positive patients) and the treatment they receive $T_{ij}$ (zero for the baseline treatment $k$ and one for the active treatment arm $l$ that are specific to study $j$);

$$\log(\lambda_{ij}) = \mu_{-ve,j} + \beta_{ij}X_{ij} + \delta_{-ve,j,kl}T_{ij} + \Delta_{j,kl} X_{ij} * T_{ij} \qquad (2)$$

In this regression model, $\mu_{-ve,j}$ and $\delta_{-ve,j,kl}$ are the baseline treatment effect and the relative treatment effect (log hazard ratio of treatment $l$ versus $k$) respectively, in study $j$ for biomarker negative patients. For biomarker positive patients, the baseline and relative treatment effects in the study $j$ are $\mu_{+ve,j}$ and $\delta_{+ve,j,kl}$ respectively, such that:

$$\beta_{ij} = \mu_{+ve,j} - \mu_{-ve,j} \qquad (3)$$

and

$$\Delta_{j,kl} = \delta_{+ve,j,kl} - \delta_{-ve,j,kl} \qquad (4)$$

Treatments $k$ and $l$ are specific to study $j$. The relative effects are assumed exchangeable within each biomarker subgroup (and within each treatment contrast), thus

$$\delta_{-ve,j,kl} \sim N(md_{-ve,kl}, \tau^2) \text{ and } \delta_{+ve,j,kl} \sim N(md_{+ve,kl}, \tau^2).$$

As in standard NMA, the mean effects $md_{-ve,kl}$ and $md_{+ve,kl}$ within each treatment contrast $l$ versus $k$ are assumed to satisfy the consistency assumption, namely

$$md_{-ve,kl} = d_{-ve,l} - d_{-ve,k}$$

$$md_{+ve,kl} = d_{+ve,l} - d_{+ve,k}$$

for each biomarker subgroup. The parameters $d_{-ve,k}, d_{-ve,l}, d_{+ve,k}, d_{+ve,l}$ are so called basic parameters, specific to the biomarker group, representing the effect of treatments $k$ or $l$ relative to the reference treatment in the network numbered as 1 and $d_{-ve,1}, d_{+ve,1} = 0$. In our implementation, prior distributions placed on the parameters specific to this part of the model are:

$$\gamma_j \sim Gamma\,(1,0.01),$$

$$\mu_{+ve,j} \sim N(0,100),\ \mu_{-ve,j} \sim N(0,100).$$

### 3.3.2. Part II: NMA model for AD studies

To allow for the assumption of normality of the treatment effects, the hazard ratios or median survival times (assuming an exponential distribution) were converted to the log hazard ratios $(y_{j,kl})$ which were then used in the model along with the corresponding standard deviations $(\sigma_j)$. We incorporated the information of the proportion of patients that are biomarker positive through the network meta-regression. The normally distributed treatment effect $(y_{j,kl})$ for each study $j$ is an estimate of a true treatment effect $\delta_{j,kl}$

$$y_{j,kl} \sim N(\delta_{j,kl}, \sigma_j^2)$$

between two treatment arms comparing treatments $k$ and $l$ ($k \neq l$ and $k, l = 1, \ldots, n_t$, with $n_t$ – number of treatments in the network). The true effects are assumed to follow a common distribution within each treatment contrast $kl$

$$\delta_{j,kl} \sim N(md_{j,kl}, \tau^2)$$

with

$$md_{j,kl} = d_{-ve,l} - d_{-ve,k} + (\bar{\beta}_l - \bar{\beta}_k) * ppos_j,$$

The basic parameters $d_{-ve,k}$ and $d_{-ve,l}$ correspond to the effects of treatments $k$ and $l$ in the biomarker negative group, as in the first part of the model for IPD studies. The parameters $\bar{\beta}_l$ and $\bar{\beta}_k$ are the study-level meta-regression coefficients corresponding to the proportion of biomarker positive participants in study $j$, denoted $ppos_j$. This results in the basic parameters for the biomarker positive group being given by;

$$d_{+ve,k} = d_{-ve,k} + \bar{\beta}_k.$$

$$d_{+ve,l} = d_{-ve,l} + \bar{\beta}_l.$$

Prior distributions are placed on the parameters specific to this part of the model:

$$\bar{\beta}_k, \bar{\beta}_k \sim N(0,100)$$

### 3.3.3. Part III: Combination of IPD and AD

We combined IPD and AD models via the shared basic parameters $d_{-ve,l}$, $d_{+ve,l}$, $d_{-ve,k}$, $d_{+ve,k}$, which are informed by both sets of studies. We place prior distributions on the parameters common to both above parts of the model:

$$d_{-ve,k}, d_{-ve,l} \sim N(0, 100)$$

$$\tau \sim U(0,2).$$

### 3.4. Application to illustrative examples

To explore the added value of the IPD, we analyse the data in our illustrative examples in stages using three models. Model 1 is the NMR model described in Section 3.3.2 (with additional prior distributions listed in Section 3.3.3. Model 2 is two-stage IPD NMA, where in stage one we analyse IPD for each study independently and use the resulting log HRs and corresponding standard deviations, together with those from AD studies, as inputs to NMR model 1 [19]. When analysing IPD for each study $j$, we use the Weibull model described in equations (1)-(4) and by placing prior distributions

$$\delta_{-ve,j,kl} \sim N(0, 100) \; and \; \delta_{-ve,j.kl} \sim N(0,100)$$

in addition to those for $\mu_{+ve,j}$, $\mu_{-ve,j}$ and $\gamma_j$.

Model 3 is a one-stage NMA, as described in Section 3.3, combining IPD and AD from all available studies.

## 4. Results

### 4.1. Breast cancer case study:

Target trial emulation in the breast cancer case study resulted in ten target trials; five comparing taxanes with chemotherapy and five comparing taxanes in combination with chemotherapy with chemotherapy alone. The baseline characteristics of patients in each emulated trial are presented in Table 1 of the Appendix. The data from target trials (described in section 2.2) were combined with AD level data from the RCTs (described in the Appendix), resulting in a network of 23 trials. All the IPD and AD studies have mixed populations of HR+ve and HR-ve patients. The network diagram of the included studies is presented in Figure 1. The solid lines illustrate treatments that have been directly compared (i.e., estimated direct effects) and the dashed line corresponds to indirect effect.

**Figure 1**: Network diagram of RCTs (marked as AD) and target trials (marked as IPD) for overall survival in the Breast Cancer case study

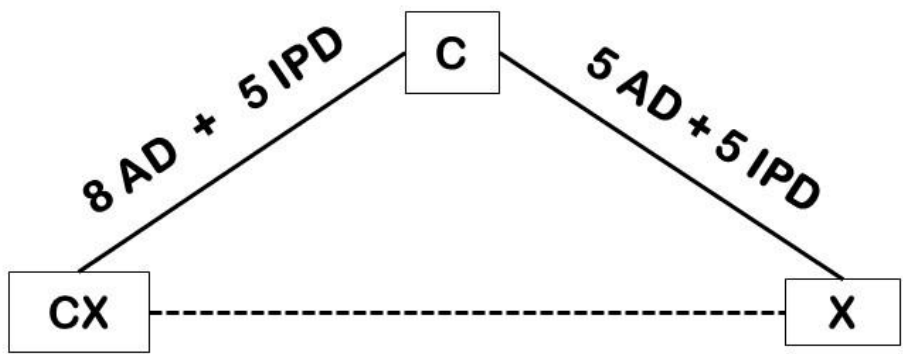

The results of the NMA for the breast cancer case study are reported in Figure 2a for the hormone receptor positive patients and Figure 2b for the hormone receptor negative patients. The hazard ratios correspond to the analyses of AD from RCTs alone (Model 1), two stage IPD meta-analysis combining data form RCTs and target trials (Model 2) and one-stage IPD (Model 3). Compared to the analysis of AD from RCTs alone, the addition of data from EHR resulted in reduced uncertainty, particularly for the hormone receptor positive patients; for example, for CX vs X the HR was 1.10 (95%CrI: 0.62, 1.84) using RCT data alone and 1.04 (95%CrI: 0.69, 1.56) when including data from target trials, which corresponded to 28.7% reduction in the width of the CrI. However, there was further substantial reduction in uncertainty, of 49% in the width of the CrI, when including data from the target trials at IPD level; to HR= 0.99 (95%CrI: 0.67, 1.29), where there was direct information on hormone receptor status of individual patients. Inclusion of target trial data at IPD level reduced uncertainty by 29% of the width of the CrI compared to the interval obtained when including target trial data in two-stage approach, where it enters NMA at AD level with the information on the biomarker status included as the proportion.

For the hormone receptor negative patients, there was a smaller reduction in uncertainty when including target trial data, compared with the results of RCT data alone; for example, for CX vs X HR=0.93 (95%CrI: 0.35, 2.01) form RCTs alone and 0.95 (95%CrI: 0.41, 1.85) from the two-stage inclusion of target trial data (Model 2), corresponding to 13.3% reduction in the width of the interval. But, as for HR+ve patients, there was a substantial reduction in uncertainty when introducing IPD form target trials (Model 3), with HR=1.00 (95%CrI: 0.65, 1.76) corresponding to 33% reduction in the width of the interval compared to RCT data alone and 29.7% reduction compared with including target trial data in the two-stage approach. However, the treatment effect estimates for all three treatment contrasts were similar for the HR+ve and HR-ve patients regardless of modelling strategy.

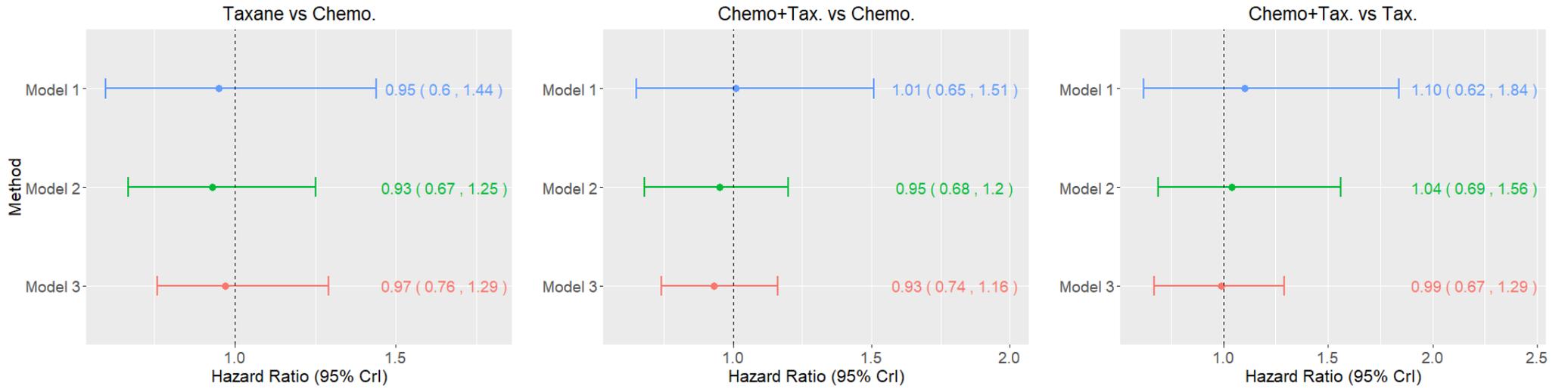

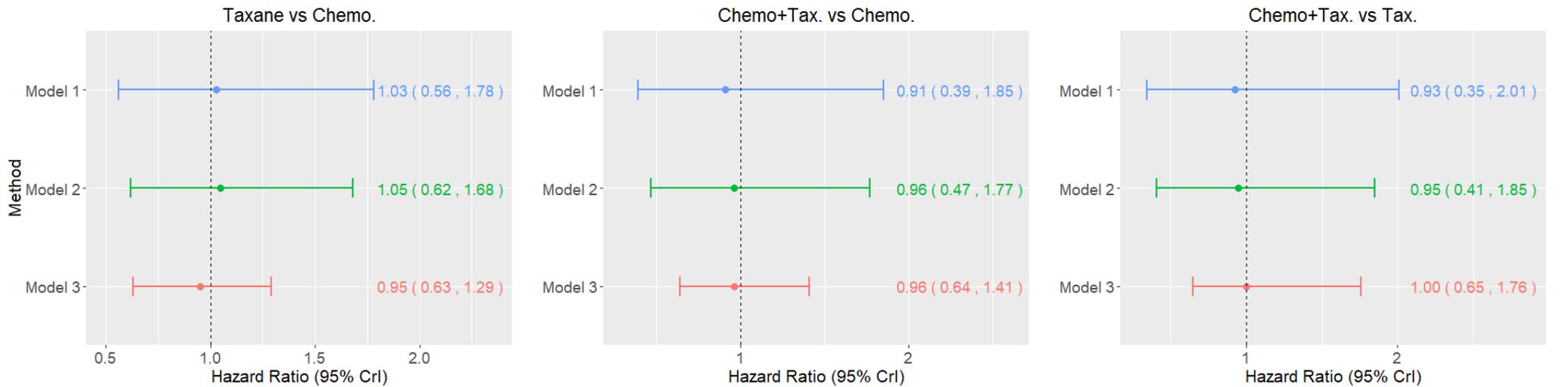

**Figure 2:** Treatment effect estimates for overall survival in breast cancer example; (a) for the hormone receptor positive patients, and (b) for the hormone receptor negative patients. Model 1: NMR of RCT data [Blue], Model 2: two-stage IPD NMA [Green], and Model 3: one-stage IPD NMA [Red].

### 4.2. Colorectal cancer example:

IPD were obtained from 10 RCTs comparing EGRF inhibitors in combination with chemotherapy with chemotherapy alone, reporting Kaplan-Meier curves for subgroups of patients with KRAS wild type and KRAS mutant status. Four of these trials included KRAS wild type patients alone. The data were combined with AD level data from 5 RCTs comparing VEGF+C with C alone, described in Section 2.1.1 to form a network of 15 studies. Network diagram for this case study is shown in Figure 3.

Figure 3: Network plot of RCTs for overall survival in the colorectal cancer case study

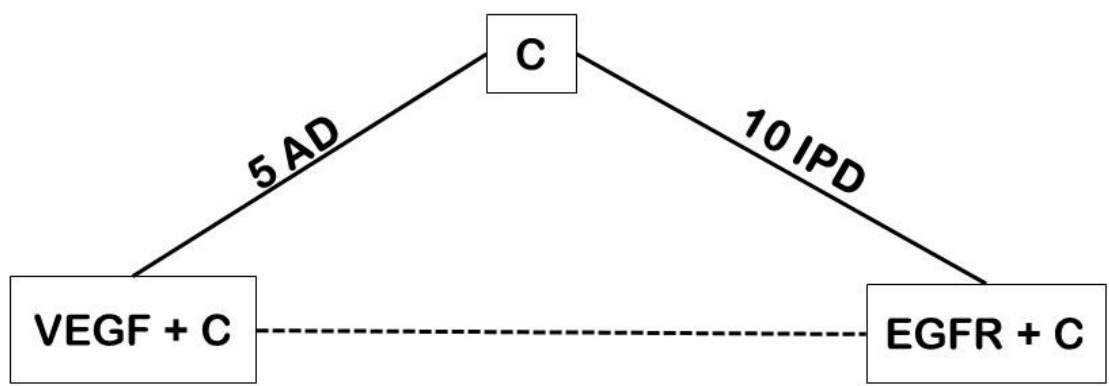

The results NMA in the colorectal cancer example are presented in Figure 4a for KRAS wild type patients and Figure 4b for patients with KRAS mutant status (MT). In contrast to the breast cancer example, where for each direct treatment contrast there was a mixture of AD and IPD, here for each treatment contrast only either IPD or AD were available. Information on the treatment comparison EGFR+C vs C is available from IPD only and information on the treatment comparison VEGF+C vs C is available from AD only. Since the IPD were available from the RCTs (via the digitised curves), we assume the data from these studies would have been available at least at the aggregate level, therefore for this case study, we only present the results of models 2 and 3.

There was meaningful positive treatment effect of EFGR+C compared to C for the KRAS WT patients with HR = 0.86 (95% CrI: 0.74, 0.98), but not for the KRAS MT patients. Including the data from EGRF+C vs C trial at IPD level resulted in 8% reduction in uncertainty around the effectiveness estimate in KRAS WT group; to HR = 0.86 (95% CrI: 0.74, 0.96) and 37% reduction for KRAS MT patients; from HR=1.07 (0.82, 1.42) to 1.04 (0.87, 1.25).

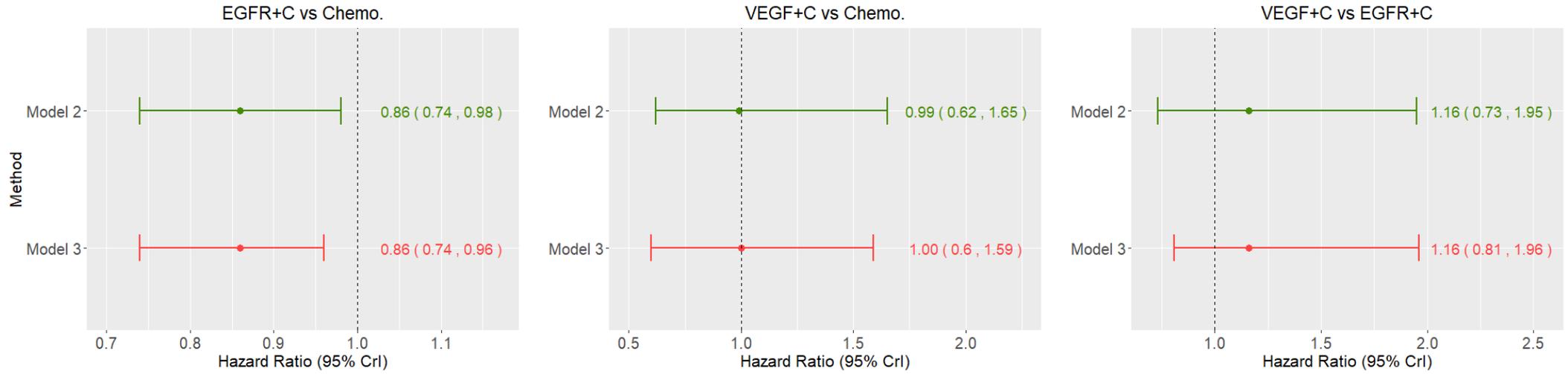

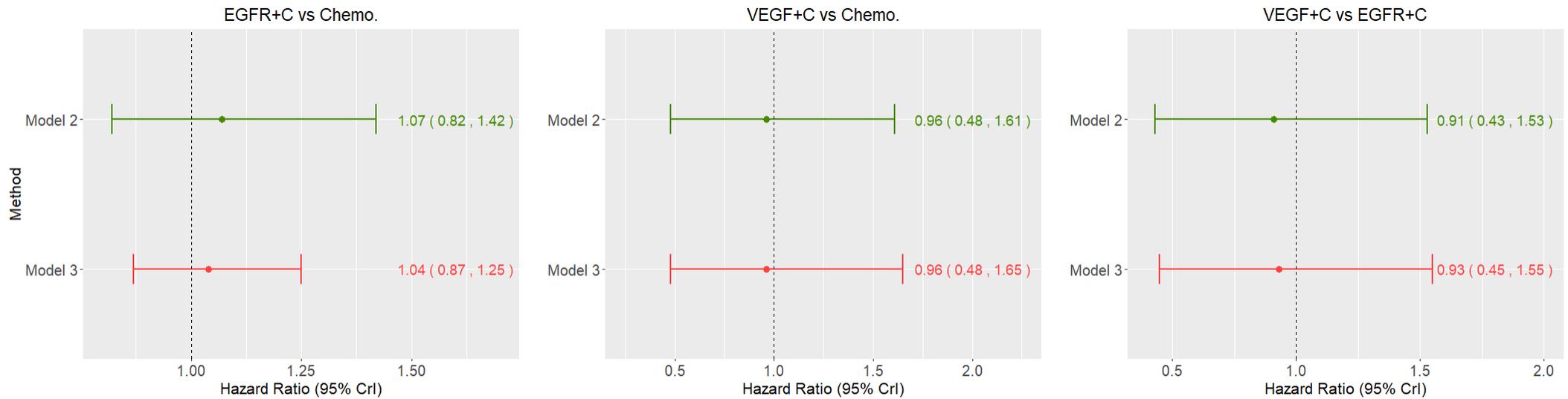

**Figure 4:** Treatment effect estimates for overall survival in the colorectal cancer case study: (a) for KRAS WT biomarker subgroup, and (b) for KRAS MT biomarker subgroup. Model 2: two-stage IPD NMA [Green], and Model 3: one-stage IPD NMA [Red].

## 5. Discussion

We developed an IPD (combined with AD) network meta regression model for estimation of treatment effects in subgroups of patients defined by a biomarker status. When subgroup analyses are not reported by RCTs, IPD are required to disentangle information on treatment effectiveness that may depend on the biomarker status. Data at individual patient level are rarely available from RCTs. We demonstrate two approaches to generating and incorporating IPD; by using EHRs or digitising Kaplan-Meier curves from RCTs that report such curves for each biomarker group. These approaches typically result in obtaining IPD for either additional studies (in our case target trials from EHRs) or a subset of RCTs (those reporting Kaplan-Meier curves). In our one-stage approach to IPD NMR we, therefore, allow for IPD to be combined with AD. The method was developed in a Bayesian framework to model time-to-event outcomes such as OS or progression-free survival.

Inclusion of data from EHR resulted in improved precision (in terms of the width of 95% CrIs) of treatment effect estimates in the breast cancer case study, even when using such data at aggregate level in the two-stage approach. This improvement was even more pronounced when including the data at the IPD level where more detailed information about the biomarker status was available. This was also the case for colorectal cancer case study, in particular for the effect of EGFR+C versus C in KRAS mutant patients. The reduction in uncertainty for the KRAS wild type patients was less noticeable, possibly since four out of ten RCTs for EGFR+C versus C were in KRAS wild type patients alone, resulting in already precise treatment effect estimate when analysing data at aggregate level.

We have demonstrated added value of IPD, that can be obtained by emulating trials from EHR or by the digitization of KM curves. The point estimates from the one-stage NMA and two – stage NMR were similar, supporting previous similar findings [20]. Estimates obtained from the colorectal cancer example are clinically meaningful as they support prior research showing that EGFR therapies are effective in patients in the KRAS wild type patients but not in patients KRAS mutant biomarker status [21]. However, use of digitized Kaplan-Meier curves as a source of IPD may mean that there could be some discrepancies with the original data. Another limitation of this research is that the IPD in the breast cancer case study were emulated from a synthetic dataset (Simulacrum) based on the Systemic Anti-Cancer Treatment (SACT) dataset rather than from the original SACT data. We used the data to illustrate our methodology, but the estimates obtained may not be meaningful from the clinical point of view.

In our method, we used a Weibull proportional hazards regression model for our time-to-event data. Other parametric methods; for example, using exponential, gamma, Gompertz or log-logistic distributions, could also be used. The model assumes proportional hazard, which is a known limitation, but this could be relaxed by adopting an Accelerated Failure Time (AFT) or flexible parametric modelling approach.

## 6. Conclusion

Inclusion of IPD in a one-stage approach to network meta-regression of time-to-event data allowed for increased precision in the estimation of treatment effects within biomarker subgroups compared to two-stage approach where only aggregate level information on the biomarker status was available. The method is particularly useful when subgroup analyses according to the biomarker status are not reported.

**Acknowledgements**

This research was funded by the Medical Research Council, Methodology Research Panel (grant no. MR/T025166/1) and partly supported by Health Data Research UK, an initiative funded by UK Research and Innovation, Department of Health and Social Care (England) and the devolved administrations, and leading medical research charities.

SB was also supported by Leicester NIHR Biomedical Research Centre (BRC). The views expressed are those of the author(s) and not necessarily those of the NIHR or the Department of Health and Social Care.

# Appendix

## 1. Emulation of target trials

Table 1 shows the emulated trials from the Simulacrum dataset.

## Table 1: Patient characteristics in the target trials

| Target trial | Treatment | Number of patients (%) | Mean age (SD) | HR+ve N (%) | HR-ve N (%) |
|---|---|---|---|---|---|
| 1 | Taxane | 67 (51) | 59.25 (17.50) | 49 (49) | 18 (58) |
|   | Chemotherapy | 65 (49) | 62.06 (14.59) | 52 (51) | 13 (42) |
| 2 | Taxane | 63 (49) | 61.87 (14.79) | 52 (51) | 11 (41) |
|   | Chemotherapy | 66 (51) | 62.93 (15.30) | 50 (49) | 16 (59) |
| 3 | Taxane | 49 (46) | 63.82 (16.73) | 39 (49) | 10 (38) |
|   | Chemotherapy | 57 (54) | 62.16 (14.69) | 41 (51) | 16 (62) |
| 4 | Taxane | 58 (53) | 63.00 (11.71) | 39 (53) | 19 (51) |
|   | Chemotherapy | 52 (47) | 62.53 (17.40) | 34 (47) | 18 (49) |
| 5 | Taxane | 59 (51) | 65.46 (14.07) | 40 (49) | 19 (58) |
|   | Chemotherapy | 56 (49) | 63.84 (14.57) | 42 (51) | 14 (42) |
| 6 | Taxane+ Chemotherapy | 81 (52) | 63.67 (14.16) | 55 (49) | 26 (59) |
|   | Chemotherapy | 75 (48) | 64.76 (15.34) | 57 (51) | 18 (41) |
| 7 | Taxane+ Chemotherapy | 69 (48) | 64.61 (13.38) | 53 (54) | 16 (35) |
|   | Chemotherapy | 76 (52) | 61.75 (15.44) | 46 (46) | 30 (65) |
| 8 | Taxane+ Chemotherapy | 71 (53) | 62.45 (12.88) | 50 (54) | 21 (53) |
|   | Chemotherapy | 62 (47) | 63.00 (16.45) | 43 (46) | 19 (47) |
| 9 | Taxane+ Chemotherapy | 60 (48) | 62.32 (15.77) | 44 (45) | 16 (59) |
|   | Chemotherapy | 65 (52) | 64.52 (13.10) | 54 (55) | 11 (41) |
| 10 | Taxane+ Chemotherapy | 63 (49) | 62.62 (15.59) | 48 (49) | 15 (48) |
|   | Chemotherapy | 66 (51) | 64.18 (15.45) | 50 (51) | 16 (52) |

SD=standard deviation; HR+ve is hormone receptor positive; HR-ve is hormone receptor negative

## 2. Evidence from RCTs with aggregate level data (AD)

RCTs were extracted from a systematic review by Ghersi et al[1], which evaluated the treatment effectiveness of taxane in breast cancer patients. Information on the overall survival measured in hazard ratio, the number of patients recruited in the study, the number of hormone receptor positive patients, and the number of hormone receptor negative patients were collected form each RCTs. Hence, RCTs that did not report information on the number of HR+ve and/or HR-ve negative patients were not included. HR+ve include patients who are progesterone and / or oestrogen positive; while HR-ve patients are patients with progesterone and oestrogen negative.

Some studies had patients whose hormone receptor status were unknown. This missingness were considered as missing at random. For patients with unknown hormone receptor status, the proportions of hormone receptor positive patients and hormone receptor negative patients were assumed to be the same as the patients with known status.

The final proportion of HR+ve patients is used in the analysis.

## Table 2: List of RCTs with aggregate dataset from the review by Ghersi et al[1]

| Author (Year) | Experimental Treatment | Reference Treatment | N | n_pos | n_neg | n_ukwn | Final n_pos | prop_pos | Overall survival |
|---|---|---|---|---|---|---|---|---|---|
| Bishop et al (1999)[2] | Paclitaxel | Cyclo +methotrexate + fluorouracil | 209 | 131 | 49 | 29 | 153 | 0.732 | 0.75(0.55,1.02) |
| Blohmer et al (2009)[3] | epirubicin + docetaxel | epirubicin + cyclo | 236 | 110 | 114 | 12 | 116 | 0.491 | 0.66(0.46,0.95) |
| Bonneterre et al (2004)[4] | epirubicin + docetaxel | fluorouracil + Epirubin + cyclo | 142 | 96 | 46 | 0 | 96 | 0.676 | 0.82(0.52,1.30) |
| Bontenbal et al (2006)[5] | doxorubicin + docetaxel | Fluorouracil + Doxorubicin+cylco | 216 | 145 | 61 | 10 | 153 | 0.708 | 0.7(o.52,0.94) |
| Campone et al (2013)[6] | docetaxel + capecitabine | Vinorelbine + capecitabine | 136 | 101 | 28 | 7 | 107 | 0.786 | 0.92(0.76,1.29) |
| Sledge et al(2003)[7] | doxorubicin + paclitaxel | doxorubicin | 454 | 203 | 117 | 134 | 289 | 0.636 | 0.98(0.77,1.36) |
| Sledge et al(2003)[7] | Paclitaxel | doxorubicin | 453 | 209 | 119 | 125 | 289 | 0.637 | 0.92(0.72,1.18) |
| Paridaens et al (2000)[8] | paclitaxel | doxorubicin | 331 | 85 | 128 | 118 | 133 | 0.401 | 1.12(0.87,1.43) |
| Biganzoli et al (2002)[9] | doxorubicin + Paclitaxel | doxorubicin + cyclo | 275 | 182 | 87 | 6 | 187 | 0.68 | 1.11(0.83,1.49) |
| Lyman et al (2004)[10] | doxorubicin + paclitaxel | doxorubicin +cyclo | 91 | 47 | 32 | 12 | 55 | 0.604 | 2.07(1.29,3.33) |
| O Shaughnessy et al (2002)[11] | Capecitabine + docetaxel | Docetaxel | 511 | 207 | 159 | 145 | 290 | 0.567 | 0.78(0.63,0.95) |
| Sjostorm et al (1999)[12] | docetaxel | methotrexate + fluorouracil | 282 | 93 | 118 | 71 | 125 | 0.443 | 0.99(0.75,1.29) |
| Sparano et al (2009)[13] | Pegylated Liposomal doxorubicin + docetaxel | docetaxel | 751 | 335 | 193 | 223 | 477 | 0.635 | 1.02(0.86,1.22) |
| Icli et al (2005)[14] | paclitaxel | cisplatin+ etoposide | 193 | 60 | 37 | 96 | 120 | 0.622 | 1.37(1.01,1.87) |
| Wang et al (2015)[15] | capecitabine + docetaxel | Vinorelbine + Capecitabine | 206 | 125 | 81 | 0 | 125 | 0.607 | 1.65(1.18,2.30) |
| Xu et al (2011) [16] | Gemcitabine + Cisplatin | Gemcitabine + paclitaxel | 99 | 47 | 20 | 32 | 70 | 0.707 | 1.49(0.82,2.73) |
| Xu et al (2011) [16] | Gemcitabine + carboplatin | Gemcitabine + paclitaxel | 96 | 47 | 20 | 29 | 68 | 0.708 | 0.81(0.44,1.50) |
| Yardley et al (2009)[17] | docetaxel | liposomal doxorubicin | 102 | 72 | 30 | 0 | 72 | 0.706 | 1.22(0.81,1.85) |

N is the total number of patients in the study; n_pos is the number of hormone receptor positive patients; n_neg is the total number of hormone receptor negative patient; n_ukwn is total number of patients with unknown hormone receptor status, final_npos is the estimated number of hormone receptor patients (see calculation of how this is obtained below); prop_pos is the proportion of hormone receptor positive.

**Table 3: List of RCTs with aggregate dataset from the review by Poad et al [18] used in the colorectal cancer case study.**

| Author (Year) | Reference Treatment | Experimental Treatment | N | Overall Survival | Prop_pos |
|---|---|---|---|---|---|
| Tebbutt et al [19] (2010) | Chemotherapy | VEGF + Chemo | 471 | 0.93 (0.75, 1.16) | 0.7134 |
| Tabernero et al [20] (2015) | Chemotherapy | VEGF + Chemo | 1072 | 0.85 (0.74, 0.98) | 0.5056 |
| Tabernero et al (2013) | Chemotherapy | VEGF + Chemo | 198 | 1.13 (0.79, 1.61) | 0.5353 |
| Hoff et al (2012) | Chemotherapy | VEGF + Chemo | 860 | 0.94 (0.79, 1.12) | 0.5767 |
| Bennouna et al (2013) | Chemotherapy | VEGF + Chemo | 819 | 0.81 (0.69, 0.94) | 0.5140 |

N is the total number of patients in the study; VEGF is vascular endothelial growth factor; chemotherapy includes and VEGF includes Bevacizumab, sorafenib and Cediranib; prop_pos is the proportion of patients with KRASWT; chemotherapy include capecitabine, FLOFOX, and FOLFIRI.

**Table 4: List of RCTs used as IPD in the colorectal cancer case study**

| Author (Year) | Reference Treatment | Experimental Treatment | Overall Survival |
|---|---|---|---|
| Ciardiello et al[21] (2016) | Chemotherapy | EGFR + Chemo | YES |
| Ye at al [22](2013) | Chemotherapy | EGFR + Chemo | YES |
| VanCutsem et al [23](2009) | Chemotherapy | EGFR + Chemo | YES |
| Seymour et al[24] (2013) | Chemotherapy | EGFR + Chemo | YES |
| Qin et al[25] (2018) | Chemotherapy | EGFR + Chemo | YES |
| Peeters et al [26](2010) | Chemotherapy | EGFR + Chemo | YES |
| Peeters et al [27](2014) | Chemotherapy | EGFR + Chemo | YES |
| Guren et al [28](2017) | Chemotherapy | EGFR + Chemo | YES |
| Douillard et al[29] (2014) | Chemotherapy | EGFR + Chemo | YES |
| Bokemeyer et al [30](2009) | Chemotherapy | EGFR + Chemo | YES |

EFGR is epidermal growth factor receptor; EGFR include Cetuximab, Panitumumab, and Gefitinib; Chemotherapy include FOLFOX, FLOFRI, Irinotecan and CAPOX.